\documentclass[
twocolumn,
aps,
prl,
reprint,
groupedaddress,
amsmath,amssymb
]{revtex4-1}
\usepackage[pdftex]{graphicx, color}
\usepackage[hidelinks]{hyperref}
\hypersetup{% hyperrefオプションリスト
 setpagesize=false,
 bookmarksnumbered=true,%
 bookmarksopen=true,%
 colorlinks=true,%
 linkcolor=blue,
 citecolor=blue,
}
\usepackage{braket}
\usepackage{times,multirow,amsfonts,bm,xspace,pifont}
\usepackage{soul}
\begin{document}
\newcommand{\rr}{{\bm r}}
\newcommand{\q}{{\bm q}}
\renewcommand{\k}{{\bm k}}
\newcommand*\wien    {\textsc{wien}2k\xspace}
\newcommand*\textred[1]{\textcolor{red}{#1}}
\newcommand*\textblue[1]{\textcolor{blue}{#1}}
\newcommand*\YY[1]{\textcolor{red}{#1}}
\renewcommand*\SS[1]{\textcolor{blue}{#1}}
\newcommand*\AD[1]{\textcolor{cyan}{#1}}
\newcommand*\JI[1]{\textcolor{red}{#1}}
\newcommand{\ki}[1]{{\color{red}\st{#1}}}

% Title of paper
\title{Insulator-Metal Transition and Topological Superconductivity in \texorpdfstring{UTe$_2$}{UTe2} \texorpdfstring{\\}{} from a First-Principles Calculation}

\author{Jun Ishizuka}
% \email[]{ishizuka.jun.8c@kyoto-u.ac.jp}
\affiliation{Department of Physics, Graduate School of Science, Kyoto University, Kyoto 606-8502, Japan}

\author{Shuntaro Sumita}
\affiliation{Department of Physics, Graduate School of Science, Kyoto University, Kyoto 606-8502, Japan}

\author{Akito Daido}
\affiliation{Department of Physics, Graduate School of Science, Kyoto University, Kyoto 606-8502, Japan}

\author{Youichi Yanase}
% \email[]{yanase@scphys.kyoto-u.ac.jp}
\affiliation{Department of Physics, Graduate School of Science, Kyoto University, Kyoto 606-8502, Japan}

\date{\today}

\begin{abstract}
% insert abstract here
We theoretically study superconductivity in UTe$_2$, which is a recently discovered strong candidate for an odd-parity spin-triplet superconductor. Theoretical studies for this compound faced difficulty because first-principles calculations predict an insulating electronic state, incompatible with superconducting instability. To overcome this problem, we take into account electron correlation effects by a GGA$+U$ method and show the insulator-metal transition by Coulomb interaction. Using Fermi surfaces obtained as a function of $U$, we clarify topological properties of possible superconducting states. Fermi surface formulas for the three-dimensional winding number and three two-dimensional $\mathbb{Z}_2$ numbers indicate topological superconductivity at an intermediate $U$ for all the odd-parity pairing symmetry in the $Immm$ space group. Symmetry and topology of superconducting gap nodes are analyzed and the gap structure of UTe$_2$ is predicted. Topologically protected low-energy excitations are highlighted, and experiments by bulk and surface probes are proposed to link Fermi surfaces and pairing symmetry. Based on the results, we also discuss multiple superconducting phases under magnetic fields, which were implied by recent experiments. 
\end{abstract}

\maketitle

%\textit{Introduction} --- 

A recent discovery of superconductivity in UTe$_2$ \cite{Ran_UTe2_2019} is attracting much attention.
Distinct differences of UTe$_2$ from other uranium-based ferromagnetic superconductors~\cite{Aoki_review,Ishida_review} are a rather high superconducting transition temperature $T_c\sim1.6$ K and a nonmagnetic behavior down to $25$ mK \cite{Sundar_UTe2}.
Uniform magnetic susceptibility, magnetization, NMR Knight shift, and $1/T_1T$ support the proximity of metallic ferromagnetic quantum criticality \cite{Ran_UTe2_2019, Aoki_UTe2, Tokunaga_UTe2}.
An extremely large upper critical field and re-entrant superconductivity have been observed by high-field experiments \cite{Ran_UTe2_2019, Ran_UTe2_H-T, Knebel_UTe2}.
These observations strongly suggest odd-parity superconductivity induced by ferromagnetic fluctuations.
A large specific heat coefficient $\gamma=117$ mJ K$^{-2}$mol$^{-1}$ indicates itinerant heavy $f$ electrons \cite{Aoki_UTe2, Knafo_UTe2, A.Miyake_UTe2}.
A large residual value of $\gamma$ in the superconducting state \cite{Ran_UTe2_2019,Aoki_UTe2} suggests a time-reversal symmetry breaking nonunitary pairing, which is known to exist in ferromagnetic superconducting states. 
However, a direct transition from a normal to a nonunitary superconducting state is prohibited in the presence of spin-orbit coupling by symmetry because the orthorhombic $D_{2h}$ point group of UTe$_2$ includes only one-dimensional (1D) representations.
Experimental studies examining this issue are in progress. 

Identifying the topological nature of quantum states has been one of the central issues in modern condensed matter physics.
Because odd-parity superconductors are a strong candidate of topological superconductors accompanied by Majorana quasiparticles \cite{Sato2009,Sato2010,Fu2010}, many studies have been focused on the odd-parity superconductivity \cite{Qi_review,Sato_review2016, Sato_review2017}. 
%Topological superconductivity accompanied by Majorana quasiparticles attracts much attention from both fundamental and practical points of view \cite{Sato_reviw2016,Sato_review2017,Qi_review}. 
However, odd-parity superconducting materials are rare. 
Therefore, identifying topological properties of a fresh and good platform UTe$_2$ is awaited.
%UTe$_2$ is a fresh and good platform. 
A nonmagnetic behavior of UTe$_2$ enables time-reversal invariant (class DIII) topological superconductivity, and a relatively high transition temperature at ambient pressure allows many experimental tools, which were hard to use for ferromagnetic superconductors \cite{Aoki_review,Ishida_review}. 
Theoretically, it is important to specify Fermi surfaces (FSs) to identify topological superconductivity. Topological invariants depend on FSs and some of them can be obtained by FS formulas \cite{Sato2009,Sato2010,Fu2010}.

Information on FSs can also be linked to gap structures of unconventional superconductors, and therefore, it enables us to study pairing symmetry by measurements of low energy excitations \cite{Sigrist-Ueda}.
Recent progress in topological theory attached a renewed attention to the gap node. 
According to modern classification of gapless superconductors \cite{Kobayashi2014,Kobayashi2016,Kobayashi2018,Sumita2019}, all the superconducting gap nodes are topologically protected. Thus, the criterion of topological superconductivity and gap nodes based on FSs provides a prediction of bulk and surface measurements, revealing the pairing symmetry and Majorana surface states.
%In this regard, clarifying the FS is crucially important.

Theoretically, a band structure has been studied for UTe$_2$ from first-principles \cite{Aoki_UTe2}.
However, the previously obtained result shows an insulating state with a small gap of $13$ meV, which contradicts metallic behaviors in electric resistivity \cite{Ran_UTe2_2019,Aoki_UTe2}. 
On the other hand, small FSs appear in another first-principles band calculation using the relativistic linearized augmented plane wave method \cite{Fujimori_UTe2}. This is also incompatible with transport measurements \cite{Niu_UTe2} indicating a large carrier density, as well as their angle-resolved photoemission spectroscopy (ARPES) detecting large intensities around the $R$ point at the Fermi level \cite{Fujimori_UTe2}.
These discrepancies between naive band structure calculations and experiments imply that the Coulomb interaction is crucially important for UTe$_2$. 
%Recent theories have shed light on the correlated electrons compared to ARPES experiments in real materials.
%For instance, the density functional theory combined with the dynamical mean field theory (DFT+DMFT) enable us to discuss complicated electronic band structure with strong correlations, revealing the reconstruction of FSs \cite{}.
% On the other hand, for UTe$_2$, there are no convincing investigations on the Coulomb interaction. It is, therefore, important to make a physical picture of band structure calculation in the view point of electronic correlations.

In this Letter, we provide the first report of a microscopic analysis linking the electronic state and superconductivity in UTe$_2$.
We show that the insulator-metal transition is induced by Coulomb interaction. For empirically reasonable values of $U$, a metallic state is realized, and FSs promise the topological superconductivity for all possible odd-parity pairings.
The superconducting gap node ensured by crystal symmetry is predicted by the group theoretical classification combined with topological arguments. 
In addition, we discuss multiple superconducting phases under magnetic fields along the $b$ axis. A phase transition inside the superconducting phase is proposed. 
%$which may possess a nodal gap structure different from that in the zero magnetic field.

\textit{GGA+U calculation} --- 
The topology of the FS is crucially important for unconventional superconductors, in particular, for the gap structure and topological superconductivity.
%The quantum oscillation and ARPES measurements have been powerful tools in this regard.
%However, the FS in UTe$_2$ is not so clear yet.
%First, we discuss the band structure without taking into account the Coulomb interaction. 
As we introduced previously, two band structures have been reported. One is insulating \cite{Aoki_UTe2} and the other is metallic with small FSs \cite{Fujimori_UTe2}. They are clearly contradicting each other and incompatible with experiments.
We therefore carry out the density functional theory (DFT) electronic structure calculations in the paramagnetic state using the \wien package \cite{blaha_2}.
We use the relativistic full-potential linearized augmented plane wave + local orbitals method within the generalized gradient approximation (GGA).
In addition to the DFT calculation providing a noninteracting band structure, we introduce the correlation effect of $f$ electrons by the GGA$+U$ method \cite{Anisimov_LDA+U_HMF}.
Details of our band calculations are given in the Supplemental Material \cite{suppl}.

\begin{figure}[ht]
\begin{center}
\includegraphics[width=84mm]{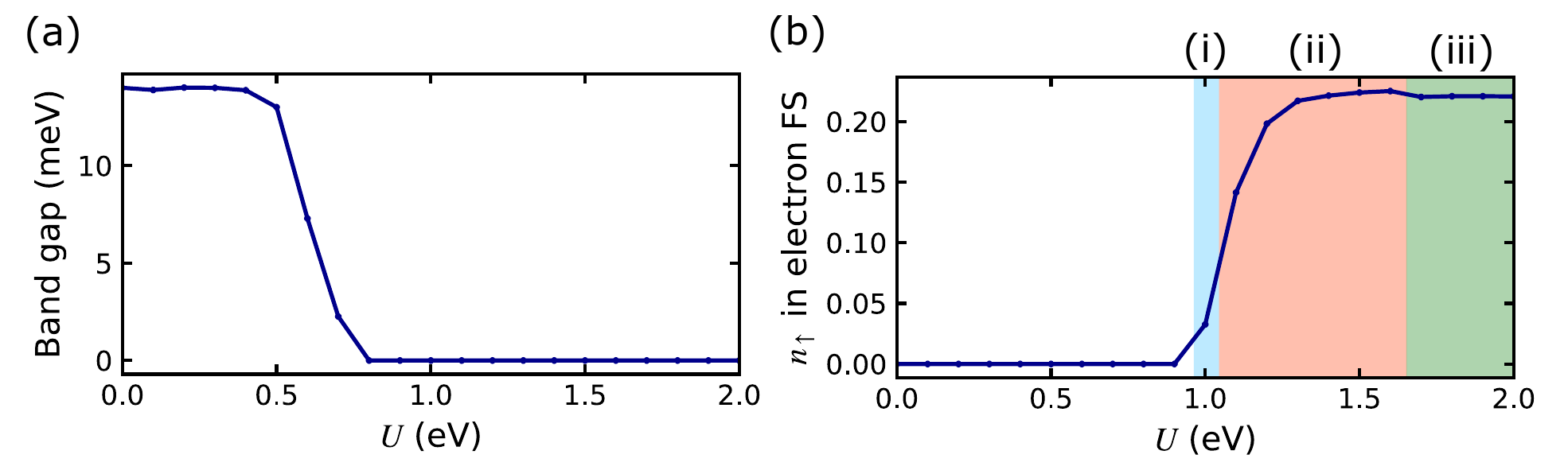}	
\caption{
Coulomb interaction dependence of (a) the band gap at the Fermi level and (b) the electron number $n$ per spin in electron FS. Insulator-metal transition occurs at $U=1.0$ eV. Metallic states with different topology of FSs are labeled by (i)-(iii).
\label{fig:gap_electron_number}}
\end{center}
\end{figure}

\begin{figure}[ht]
\begin{center}
\includegraphics[width=74mm]{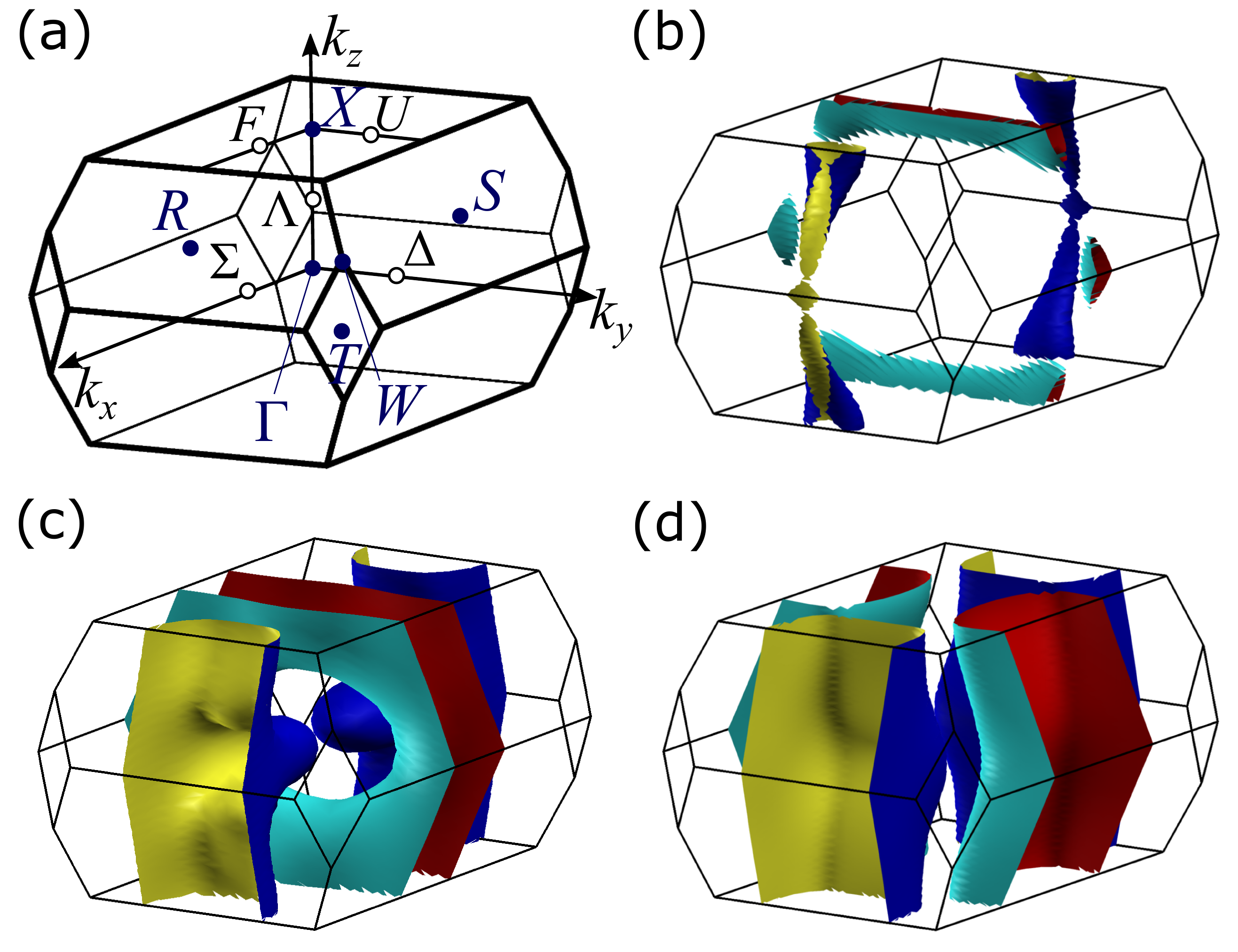}	
\caption{
(a) First BZ and symmetry points. (b)-(d) FSs of UTe$_2$ by GGA$+U$ for (b) $U=1.0$ eV [region (i)], (c) $U=1.1$ eV [region (ii)], and (d) $U=2.0$ eV [region (iii)]. The electron sheet (cyan and red colors) and the hole sheet (blue and yellow colors) \cite{suppl}.
\label{fig:FS}}
\end{center}
\end{figure}

The numerical results are given in Figs.~\ref{fig:gap_electron_number}-\ref{fig:FS}.
The DFT band structure is insulating, and the band gap is $14$ meV. 
Thus, the results are consistent with Ref.~\cite{Aoki_UTe2}. 
% Moreover, this material is identified as a strong topological insulator \cite{}.
On the other hand, the GGA$+U$ calculation shows the closing of band gap (Fig.~\ref{fig:gap_electron_number}), and we observe metallic FSs for $U>1.0$ eV.
It turns out that the correlation effect of $f$ electrons causes an insulator-metal transition. 
A moderate value of $U$ in the GGA$+U$ calculation may be reliable at low temperatures, where the itinerant $f$ electrons form a heavy fermion state consistent with specific heat measurements \cite{Ran_UTe2_2019, Aoki_UTe2}.
A larger $U$ may be appropriate above the Kondo temperature, where the $f$ electrons are localized.
Although we cannot determine the value of $U$ in the framework of the GGA$+U$ method, we can deduce which FS is more appropriate using the comparison with future experiments such as ARPES or quantum oscillations.
Furthermore, the GGA$+U$ calculation should be compared with other methods such as GGA+DMFT, which is left as a future work.

The obtained FSs are illustrated in Figs.~\ref{fig:FS}(b)-\ref{fig:FS}(d), each of which shows different topologies labeled by (i)-(iii) in Fig.~\ref{fig:gap_electron_number}(b).
For $U=1.0$ eV, a tiny electron sheet appears at the Brillouin zone (BZ) boundary around the $X$ point and a tiny hole sheet around the $R$ point.
The FSs dramatically increase their volume by an increase of $U$, involving a topological Lifshitz transition from (i) to (ii).
For $U=1.1$ eV, there appears a ringlike sheet with a heavy effective mass dominated by $j_z=\pm5/2$ and $\pm1/2$ components \cite{suppl}. We also see a cylindrical sheet dominated by $j_z=\pm5/2$ and $\pm3/2$ components.
The ringlike sheet changes to a two-dimensional (2D) cylindrical sheet at $U=2.0$ eV, which is shown in Fig.~\ref{fig:FS}(d), with a topological transition at $U \simeq 1.6$ eV.
These Fermi sheets consist of $j_z=\pm5/2$ and $\pm3/2$ components mixed with $d_{3z^2-r^2}$ and $p_y$ orbitals having a light effective mass \cite{suppl}.
The electron sheet has a large carrier density $n\sim0.2$ per spin which is compensated by the hole sheet. Thus, we see large FSs occupying  $40\%$ of the BZ, in accordance with transport measurements \cite{Niu_UTe2}.
On the other hand, low carrier density shown in Fig.~\ref{fig:FS}(b) %, the DFT band structures shifting the $5f$ level \cite{Harima_UTe2}, 
or that obtained by a slight carrier doping to the DFT band structures is incompatible with experiments for UTe$_2$.
We also confirmed the insulator-metal transition by the density of states \cite{suppl}.
Then, we notice that the incipient hybridization gap is shifted upwards in energy by $10$ meV ($U=1.0$ eV).
The $f$-electron states with $j=5/2$ multiplet are dominant around the Fermi level.

\textit{Topological superconductivity} --- 
Here, we discuss topological properties of UTe$_2$, assuming odd-parity superconductivity.
% As discussed previously, this is a natural assumption, because spin-triplet superconductivity is strongly suggested from the huge upper critical fields[@@].
We also assume that time-reversal symmetry is preserved in the superconducting phase since this is natural from the group-theoretical perspective as discussed previously.
%\textcolor{blue}{
Thus, we focus on the zero-magnetic-field phase and the putative nonunitary pairing state \cite{Ran_UTe2_2019} is beyond our scope.
% We note that origin of the residual density of states and the possibility of nonunitary paring is still under debate (cf. contradicting with thermal conductivity measurements by Izawa group)}.

Topology of odd-parity superconductors is closely related to the topology of FSs.
Actually, the parity of various topological invariants is determined by the occupation number at high-symmetry points in the BZ \cite{Sato2009,Sato2010,Fu2010}.
For example, we can identify the parity of the three-dimensional (3D) winding number $\omega$ by the following formula:
\begin{equation}
\omega=\frac{1}{2}\sum_{K_i} n(K_i)\quad(\mathrm{mod}\ 2).
\end{equation}
Here, $K_i$ runs over eight time-reversal invariant momenta (TRIM) in the 3D BZ, and the occupation number $n(K_i)$ is an even integer due to Kramers degeneracy.
Similar formulas are also known for 1D and 2D $\mathbb{Z}_2$ invariants \cite{Sato2009,Sato2010,Fu2010}.

In this Letter, we concentrate on topological invariants related to the $(100)$, $(010)$, and $(001)$ surface states.
It should be noted that the size of the primitive cell is doubled because the translation symmetry on these surfaces is compatible with the doubled unit cell. 
For this reason, we apply the formulas to the folded BZ, instead of the original one.
To be specific, the former and the latter BZ correspond to the unit cell formed by $\{a\hat{x},b\hat{y},c\hat{z}\}$ and $\{(-a\hat{x}+b\hat{y}+c\hat{z})/2,(a\hat{x}-b\hat{y}+c\hat{z})/2,(a\hat{x}+b\hat{y}-c\hat{z})/2\}$, respectively.
% For this reason, we apply the formulas to the folded BZ (corresponding to the enlarged unit cell formed by $\{a\hat{x},b\hat{y},c\hat{z}\}$), instead of the original one (corresponding to the primitive unit cell formed by $\{a\hat{x},b\hat{y},(a\hat{x}+b\hat{y}+c\hat{z})/2\}$).
%\textcolor{blue}{We follow the convention of [Ono]}, and
The correspondence between high-symmetry points in the two BZs is shown in Table~\ref{tab:folding}.

We obtained from the GGA$+U$ results the occupation number at TRIM and corresponding topological invariants (Table~\ref{tab:OccNum}).
Here, $\nu_1$, $\nu_2$, and $\nu_3$ are the $\mathbb{Z}_2$ invariants defined on the $k_x=0$, $k_y=0$, and $k_z=0$ planes, respectively.
$\mathbb{Z}_2$ invariants defined on the other time-reversal invariant planes are obtained from $(\omega,\nu_1,\nu_2,\nu_3)$, as is the case for topological insulators, and they are trivial in our results.
%1D $\mathbb{Z}_2$ invariants defined on each $k_x$, $k_y$, and $k_z$ axis are nontrivial for FSs (i) and (ii), while those defined elsewhere and those for FSs (iii) are trivial.

According to Table~\ref{tab:OccNum}, superconductivity in UTe$_2$ is topologically nontrivial for moderate values of $U$ in the regions (i) and (ii) when the bulk state is gapped. Majorana states appear on the $(100)$, $(010)$, and $(001)$ surfaces.
This is one of the central results of this Letter. 

Here we comment on the effect of excitation nodes on the topological superconductivity. 
Although the winding number $\omega$ is ill-defined in the gapless states, some of the 2D $\mathbb{Z}_2$ invariants $\nu_i$ may still be well defined and the corresponding surface Majorana states may appear.
Well-defined topological invariants and surfaces hosting Majorana states are summarized in Table~\ref{tab:surface} for each pairing symmetry and FSs. 
For example, $B_{1u}$ superconducting state has point nodes on the $k_z$ axis, meaning $\nu_1$ and $\nu_2$ are ill-defined. However, $\nu_3$ is well defined and $(100)$ and $(010)$ surface states are robust.
Thus, topological superconductivity can be detected with scanning tunneling microscopy or ARPES for clean surfaces even in gapless superconducting states.
%Existence of Majorana surface state depends on gap structures which are discussed later. 

It should be noticed that our results do not exclude the possibility of topological superconductivity for FSs (iii): Indeed, 3D winding number $\omega \in \mathbb{Z}$ can be a finite even integer.
By symmetry, this is allowed only for the $A_u$ pairing state \cite{Yoshida2019}.
We do not discuss this case because $\omega$ depends on detailed properties of superconducting gap function. We also leave the possibility of topological crystalline superconductivity as a future issue.

%\textcolor{blue}{The effect of possible excitation nodes is discussed in the following, after their detailed classification.}

\begin{table}[tbp]
 \centering
 \caption{Correspondence between high-symmetry points in the original BZ and folded BZ.
% High symmetry points in the folded BZ are indicated by the subscript $p$ along with their position shown in the form of $(k_x,k_y,k_z)$.
}
 \label{tab:folding}
 \begin{tabular*}{1.0\columnwidth}{@{\extracolsep{\fill}}p{0mm}llp{0mm}}
   \hline\hline
   & Original & Folded & \\\hline
   & $\Gamma$, $X$ & $\Gamma_p:\,(0,0,0)$ & \\
   & $R$& $U_p:\,[(\pi/a),0,(\pi/c)]$ & \\
   & $S$& $T_p:\,[0,(\pi/b),(\pi/c)]$ & \\
   & $T$& $S_p:\,[(\pi/a),(\pi/b),0]$ & \\
   & $W$& $R_p:\,[(\pi/a),(\pi/b),(\pi/c)]$ & \\
   \hline\hline
 \end{tabular*}
\end{table}

\begin{table}[tbp]
 \centering
 \caption{Occupation number $n(K_i)$ at high symmetry points in the folded BZ and topological invariants (modulo two) corresponding to each topology of FSs.
 The values $n(K_i)-180$ are shown below, with $X_p=(\pi/a,0,0)$, $Y_p=(0,\pi/b,0)$, and $Z_p=(0,0,\pi/c)$.}
 \label{tab:OccNum}
 \begin{tabular*}{1.0\columnwidth}{@{\extracolsep{\fill}}lp{2mm}cccccccccc}
   \hline\hline
FSs & & $\Gamma_p$ & $X_p$ & $S_p$ & $Y_p$ & $Z_p$ & $U_p$ & $R_p$ & $T_p$&&$(\omega,\nu_1,\nu_2,\nu_3)$\\ \hline
  (i)   & & $6$ & $4$ & $4$ & $8$ & $4$ & $0$ & $4$ & $4$&&$(1,1,1,1)$ \\
  (ii) & & $6$ & $0$ & $4$ & $8$ & $4$ & $0$ & $4$ & $8$&&$(1,1,1,1)$ \\
  (iii) & & $4$ & $0$ & $4$ & $8$ & $4$ & $0$ & $4$ & $8$ &&$(0,0,0,0)$   \\        \hline\hline
 \end{tabular*}
\end{table}

\begin{table*}[tbp]
 \centering
 \caption{Gap structures, nontrivial topological indices, and surfaces hosting stable Majorana states for odd-parity pairing states.   % We show the two cases for the $B_{1u}$ and $B_{2u}$ states because gap structure is different between FSs (i,ii) and (iii). 
}
 \label{tab:surface}
 \begin{tabular*}{1.0\textwidth}{@{\extracolsep{\fill}}lccclccc}
   \hline\hline
   FSs(i,ii) & & & & \multicolumn{4}{l}{FSs(iii)}\\
   IR & Gap structure & Topological index & Surfaces & IR & Gap structure & Topological index & Surfaces  \\\hline
   $A_u$ & Full gap & $(\omega,\nu_1,\nu_2,\nu_3)$ & (100), (010), (001) & $A_u$ & Full gap & $\omega \in 2\mathbb{Z} $ & Unpredicted \\
   $B_{1u}$ & Point node ($\Lambda$) & $\nu_3$ & (100), (010) & $B_{1u}$ & Full gap & Trivial & None \\
   $B_{2u}$ & Point node ($\Delta$) & $\nu_2$ & (100), (001) & $B_{2u}$ & Point node ($\Delta, U$) & Trivial & None \\
   $B_{3u}$ & Point node ($\Sigma, F)$ & $\nu_1$ & (010), (001) & $B_{3u}$ & Point node ($\Sigma, F)$ & Trivial & None \\
   \hline\hline
 \end{tabular*}
\end{table*}

% \begin{table}[tbp]
%  \centering
%  \caption{Gap structures, Well-defined topological invariants, and surfaces hosting stable Majorana states for odd-parity pairing states. We show the two cases for the $B_{1u}$ and $B_{2u}$ states because gap structure is different between FSs (i,ii) and (iii). 
% }
%  \label{tab:surface}
%  \begin{tabular*}{1.0\columnwidth}{@{\extracolsep{\fill}}lccc}
%   \multicolumn{4}{l}{(a) FSs(i,ii)}\\\hline\hline
%   IR & Gap structure & Topological index & Surfaces\\\hline
%   $A_u$ & Full gap & $W,\nu_{1,2,3}=1$ & (100), (010), (001) \\
%   $B_{1u}$(i,ii) & Point node ($\Lambda$) & $\nu_3=1$ & (100), (010) \\
%   $B_{1u}$(iii) & Full gap & $W,\nu_{1,2,3}=0$ & \\
%   $B_{2u}$(i,ii) & Point node ($\Delta$) & $\nu_2=1$ & (100), (001) \\
%   %$B_{2u}$(iii) & Point node ($\Delta, U$) & $\nu_2$ & (100), (001) \\
%   $B_{2u}$(iii) & Point node ($\Delta, U$) & $\nu_2=0$ & \\
%   $B_{3u}$ & Point node ($\Sigma, F)$ & $\nu_1=1$ & (010), (001) \\
%   \hline\hline
%  \end{tabular*}
% \end{table}

\textit{Topological gap node} --- 
Now we discuss gap structures in the superconducting state of UTe$_2$, using group theory and topology.
%~\cite{Volovik1984, Volovik1985, Anderson1984, Sigrist-Ueda, Yarzhemsky1992, Yarzhemsky1998, Yarzhemsky2000, Yarzhemsky2003, Yarzhemsky2008, Norman1995, Micklitz2009, Micklitz2017_PRB, Nomoto2017, Micklitz2017_PRL, Sumita2017, Sumita2018, Kobayashi2018, Sumita2019}.
First, we consider an ordinary classification theory of the superconducting \textit{order parameter} by the crystal point group~\cite{Volovik1984, Volovik1985, Anderson1984, Sigrist-Ueda}.
At zero magnetic field, UTe$_2$ possesses $D_{2h}$ point group symmetry, in which an odd-parity order parameter is classified as one of four irreducible representations (IRs): $A_u$, $B_{1u}$, $B_{2u}$, and $B_{3u}$. Typical basis functions are shown in Table~\ref{tab:classification_OP}(a). 
We also consider magnetic fields along the $b$ axis in which UTe$_2$ shows extremely high critical fields and metamagnetic transition \cite{Ran_UTe2_H-T,Ran_UTe2_2019,Knebel_UTe2,Knafo_UTe2,A.Miyake_UTe2}. 
Then, the symmetry is reduced to $C_{2h}^y$, which has two odd-parity IRs ($A_u$ and $B_u$).
In this case, the $A_u$ and $B_{2u}$ ($B_{1u}$ and $B_{3u}$) states are not distinguished by symmetry since they are reduced to the $A_u$ ($B_u$) state. 
The correspondence is summarized in Table~\ref{tab:classification_OP}(b).

\begin{table}[tbp]
 \centering
 \caption{Classification of odd-parity superconducting order parameters for point groups (a) $D_{2h}$ and (b) $C_{2h}^y$.}
 \label{tab:classification_OP}
 \begin{tabular*}{1.0\columnwidth}{@{\extracolsep{\fill}}lp{2mm}rrrrrrrrp{2mm}c}
  \multicolumn{12}{c}{(a) $D_{2h}$ (zero magnetic field)} \\ \hline\hline
  IR & & $E$ & $C_{2z}$ & $C_{2y}$ & $C_{2x}$ & $I$ & $\sigma_z$ & $\sigma_y$ & $\sigma_x$ & & Basis functions \\ \hline
  $A_u$    & & $1$ & $1$ & $1$ & $1$ & $-1$ & $-1$ & $-1$ & $-1$ & & $k_x \hat{x}, k_y \hat{y}, k_z \hat{z}$ \\
  $B_{1u}$ & & $1$ & $1$ & $-1$ & $-1$ & $-1$ & $-1$ & $1$ & $1$ & & $k_y \hat{x}, k_x \hat{y}$              \\
  $B_{2u}$ & & $1$ & $-1$ & $1$ & $-1$ & $-1$ & $1$ & $-1$ & $1$ & & $k_x \hat{z}, k_z \hat{x}$              \\
  $B_{3u}$ & & $1$ & $-1$ & $-1$ & $1$ & $-1$ & $1$ & $1$ & $-1$ & & $k_z \hat{y}, k_y \hat{z}$              \\ \hline\hline
 \end{tabular*}
 \\[2mm]
 \begin{tabular*}{1.0\columnwidth}{@{\extracolsep{\fill}}lcp{2mm}rrrrp{2mm}c}
  \multicolumn{9}{c}{(b) $C_{2h}^y$ (magnetic field $\parallel b$)} \\ \hline\hline
  IR & $(\text{IR}) \uparrow D_{2h}$ & & $E$ & $C_{2y}$ & $I$ & $\sigma_y$ & & Basis functions \\ \hline
  $A_u$ & $A_u + B_{2u}$    & & $1$ & $1$ & $-1$ & $-1$ & & $k_x \hat{x}, k_z \hat{x}, k_y \hat{y}, k_x \hat{z}, k_z \hat{z}$ \\
  $B_u$ & $B_{1u} + B_{3u}$ & & $1$ & $-1$ & $-1$ & $1$ & & $k_y \hat{x}, k_x \hat{y}, k_z \hat{y}, k_y \hat{z}$              \\ \hline\hline
 \end{tabular*}
\end{table}

Although we can speculate gap structures from the order parameter, it is desirable to use the classification of \textit{gap structures} in terms of symmetry and topology~\cite{Yarzhemsky1992, Norman1995, Micklitz2009, Nomoto2017, Micklitz2017_PRL, Sumita2017, Sumita2018, Kobayashi2014, Kobayashi2016, Kobayashi2018, Sumita2019} because symmetry-protected gap nodes are precisely obtained.
Detailed results of the topological classification are shown in the Supplemental Material~\cite{suppl}.
Using the results, gap structures of UTe$_2$ are obtained for each pairing symmetry and FSs, as shown in Table~\ref{tab:surface}.

%Under zero magnetic field, all gap structures appearing in Table~S2(a) are consistent with the order parameter analysis in Table~\ref{tab:classification_OP}(a), since the symmorphic space group has no unusual algebraic structure even on the BZ boundary, contrary to the nonsymmorphic case~\cite{Norman1995, Micklitz2009, Micklitz2017_PRB, Nomoto2017, Micklitz2017_PRL, Sumita2017, Sumita2018, Kobayashi2018}.
Considering the FS topology in Figs.~\ref{fig:FS}(b)-\ref{fig:FS}(d), we find that UTe$_2$ is a fully gapped superconductor at zero magnetic field when the order parameter belongs to $A_u$ IR with $U > 1.0$ eV [regions (i)-(iii)], or $B_{1u}$ IR with $U > 1.6$ eV [region (iii)]. 
Otherwise the superconducting state has some point nodes, whose positions depend on pairing symmetry and they can be distinguished by experiments.
In Table~\ref{tab:thermal_conductivity}, we show expected anisotropy of thermal conductivity \cite{Joynt_UPt3}, which may determine the symmetry of superconductivity in UTe$_2$.

\begin{table}[tbp]
 \centering
 \caption{Expected anisotropy in thermal conductivity at low temperatures for each FSs and pairing symmetry. Anisotropy of fully gapped states cannot be predicted.}
 \label{tab:thermal_conductivity}
 \begin{tabular*}{1.0\columnwidth}{@{\extracolsep{\fill}}lcccc} \hline\hline
  FSs & $A_u$ & $B_{1u}$ & $B_{2u}$ & $B_{3u}$ \\ \hline
  (i) & Unpredicted & $\kappa_c > \kappa_{a, b}$       & $\kappa_b > \kappa_{a, c}$ & $\kappa_a > \kappa_{b, c}$ \\
  (ii) & Unpredicted & $\kappa_c > \kappa_{a, b}$       & $\kappa_b > \kappa_{a, c}$ & $\kappa_a > \kappa_{b, c}$ \\
  (iii) & Unpredicted &  Unpredicted & $\kappa_b > \kappa_{a, c}$ & $\kappa_a > \kappa_{b, c}$ \\ \hline\hline
 \end{tabular*}
\end{table}

% The paragraph should be in the next section ?
When the magnetic field is applied along the $b$ axis, superconducting states are classified into $A_u$ or $B_u$ IR. The $A_u$ state of $C_{2h}^y$ can be regarded as a $A_u + B_{2u}$ state, a mixed representation in $D_{2h}$, while $B_u \uparrow D_{2h}= B_{1u}+B_{3u}$. 
According to the gap classification in Table~S2(b) of Supplemental Materials~\cite{suppl}, the $A_u$ state has symmetry-protected point nodes on the $k_y$ axes while the $B_u$ state has a line node on the $k_y=0$ plane. 
These results are consistent with speculation from classification of order parameter.
Since the spin-triplet order parameter with $d$-vector parallel to the magnetic field does not cause the gap of Bogoliubov quasiparticles, we have only to consider the others. For $A_u$ IR, allowed bases are $k_x \hat{x}$, $k_z \hat{x}$, $k_x \hat{z}$, and $k_z \hat{z}$, which create point nodes.
On the other hand, the $B_u$ order parameter of $k_y \hat{x}$ and $k_y \hat{z}$ results in line nodes on the $k_y = 0$ plane.
%Note that the presence of line nodes is compatible with Blount's theorem \cite{Kobayashi2014,Blount1985} because the time-reversal symmetry is broken.
%The above descriptions are consistent with Table~S2(b).

\begin{figure}[ht]
\begin{center}
\includegraphics[width=72mm]{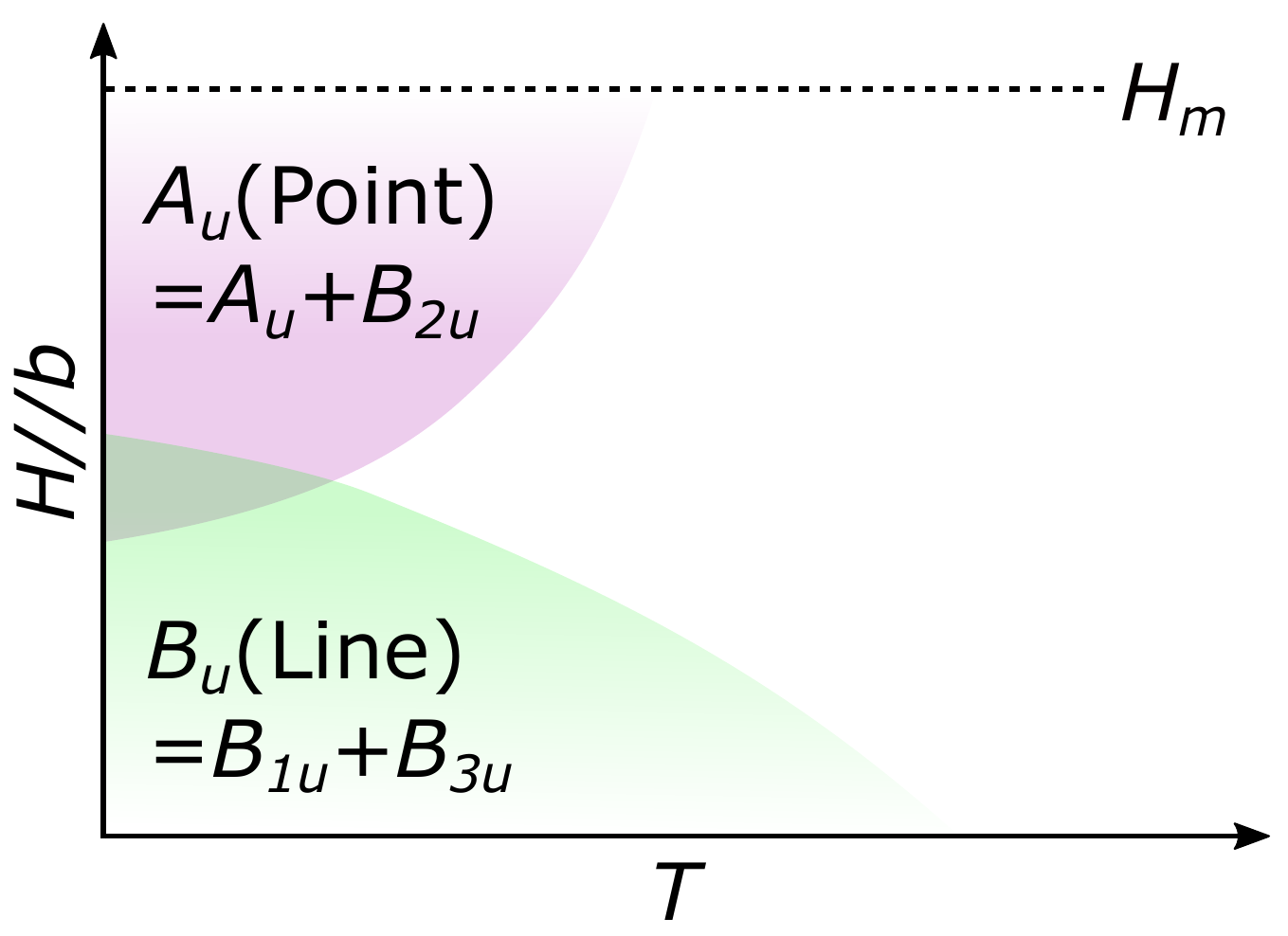}
\caption{
Schematic phase diagram in the magnetic field ($H$) and temperature ($T$) plane. $H_m$ indicates the metamagnetic transition \cite{Knafo_UTe2, A.Miyake_UTe2}. 
\label{fig:H-T_phase}}
\end{center}
\end{figure}

\textit{Multiple phases under magnetic fields} --- 
Experimental data for UTe$_2$ under magnetic fields along the $b$ axis reveal highly unusual behaviors \cite{Ran_UTe2_H-T,Ran_UTe2_2019,Knebel_UTe2}. The transition temperature shows a nonmonotonic behavior as a function of the magnetic field. It indicates the presence of two superconducting phases: The low-field and high-field phases may be distinguished by symmetry. Considering the gap structure discussed above, we propose the phase diagram in Fig.~\ref{fig:H-T_phase}. Because point-nodal superconducting states gain more condensation energy than line-nodal one, the $A_u$ state may be stable at high magnetic fields, while the $B_u = B_{1u}+B_{3u}$ state may be favored by the spin-orbit coupling at low fields. In this case, the $B_{1u}$ or $B_{3u}$ state is realized at zero magnetic field. The order parameter of both states contains $d$-vector parallel to the $b$ axis, and therefore, the Knight shift decreases below $T_{\rm c}$. This is consistent with a recent NMR experiment \cite{Nakamine_UTe2}. %\cite{Ishida_private}

Anisotropy of pairing interaction \cite{Yanase2003} is also important. 
Strongly anisotropic ferromagnetic fluctuation may favor the $d$-vector perpendicular to the easy axis.
Then, the easy axis along $a$ \cite{Ran_UTe2_2019} implies the $B_{3u}$ state at $H=0$ while the fluctuation along the $b$ axis near the metamagnetic transition \cite{Knafo_UTe2,A.Miyake_UTe2} favors the $B_{2u}$ state.
This is consistent with our proposal.  Microscopic calculations are desirable for more precise discussions and left for a future study.

\textit{Magnetism} ---
Figures~\ref{fig:FS}(c) and (d) show nesting property of the FSs,
%: The low-dimensionality originates from an U chain structure along $a$ axis and Te(2) chain along $b$ axis.
and therefore, we expect a magnetic fluctuation with a finite-$q$ nesting vector coexisting with a widely believed ferromagnetic fluctuation.
%By calculating magnetic fluctuation, we obtain a nesting property in addition to ordinarily ferromagnetic fluctuation.
Thus, it is indicated that UTe$_2$ is in the vicinity of the multiple magnetic orderings, and the long-range magnetic order is suppressed by magnetic frustration. This may explain why UTe$_2$ does not undergo magnetic order in contrast to UCoGe and URhGe. A recent first-principles study proposes a similar scenario \cite{Xu_UTe2}.

\textit{Summary and conclusion} ---
In this Letter, we theoretically investigated the electronic state and superconductivity of UTe$_2$.
Using the GGA$+U$ calculation we clarified the insulator-metal transition due to Coulomb interaction. The metallic band structure for $U>1.0$ eV is compatible with the metallic conductance and superconducting instability, indicating a crucial role of electron correlation.  
For moderate $U$, all the odd-parity superconducting states expected in UTe$_2$ were identified as time-reversal-invariant topological superconductivity. 
Superconducting gap structures under magnetic fields along the $b$ axis as well as at zero magnetic field were predicted. 
By this work, bulk and surface excitations characterizing the odd-parity superconductivity in UTe$_2$ are elucidated. 
Our results pave a way to experimentally determine symmetry of superconductivity and identify intrinsic topological superconductivity which has been rare in nature. 
%We stress that the pairing symmetry can be revealed by both bulk/surface measurements, although the mechanism of superconductivity is a challenging issue and beyond the scope of this Letter. 

After we wrote the first manuscript, we become aware of Ref.~\cite{Shick_UTe2}, which reported DFT$+U$ calculations in both paramagnetic and ferromagnetic states. At their parameters $U=J=0.51$ eV, an insulating band structure was obtained consistent with our calculations, and insulator-metal transition was not shown. 
We are also aware of an experimental paper \cite{Metz_UTe2} in which point nodes along the $a$ axis are concluded. This implies the $B_{3u}$ state at zero magnetic field consistent with our proposal in Fig.~\ref{fig:H-T_phase}.

\begin{acknowledgments}
We appreciate helpful discussions with H. Ikeda, K. Izawa, K. Ishida, S. Fujimori, D. Aoki, and H. Harima.
This work was supported by a Grant-in-Aid for Scientific Research on Innovative Areas ``J-Physics'' (Grant No.~JP15H05884) and ``Topological Materials Science'' (Grants No.~JP18H04225) from Japan Society for Promotion of Science (JSPS) and by JSPS KAKENHI (Grants No.~JP15H05745, No.~JP17H09908, No.~JP17J10588, No.~JP18H01178, and No.~JP18H05227).
%S.~S.~was supported by JSPS KAKENHI (Grants No.~JP17H09908) and A.~D.~was supported by JSPS KAKENHI (Grants No.~17J10588).
Numerical calculations have been done at the supercomputer of the ISSP in Japan.
\end{acknowledgments}

\clearpage

\renewcommand{\thesection}{S\arabic{section}}
\renewcommand{\theequation}{S\arabic{equation}}
\setcounter{equation}{0}
\renewcommand{\thefigure}{S\arabic{figure}}
\setcounter{figure}{0}
\renewcommand{\thetable}{S\arabic{table}}
\setcounter{table}{0}
\renewcommand*{\citenumfont}[1]{S#1}
\renewcommand*{\bibnumfmt}[1]{[S#1]}
\makeatletter
\c@secnumdepth = 2
\makeatother

\onecolumngrid

\begin{center}
{\large \textmd{Supplemental Materials:} \\[0.3em]
{\bfseries Insulator-Metal Transition and Topological Superconductivity in UTe$_2$ \\ from a First-Principles Calculation}}
\end{center}

\section{Details of band calculation}
We perform the DFT band structure calculation in the paramagnetic state of UTe$_2$ using the \wien code \cite{blaha_2_suppl}.
The crystallographical parameters are shown in Table \ref{tb:struct} \cite{Haneveld_UTe2_1970_suppl}.
UTe$_2$ crystallizes in an orthorhombic body-centered structure (space group No.~71, $Immm$) and the local site symmetry of uranium atoms is $C_{2v}$, lacking the inversion symmetry.
%The crystal structures are taken from Ref.~\cite{Haneveld_UTe2_1970_suppl}.
%The crystal structures of the high temperature phase are taken from Ref.~\cite{Haneveld_UTe2_1970}, but band structures are almost unchanged by that of the low-temperature phase.
The maximum reciprocal lattice vector $K_{\rm max}$ was given by $R_{\rm MT}K_{\rm max}=11.0$. The muffin-tin radii $R_{\rm MT}$ of 2.50 a.u. was chosen for U and Te atoms, and $12\times12\times12$ $k$-points sampling was used for the self-consistent calculation.
For the GGA$+U$ calculation, we set Hund's coupling $J=0$ eV.
We use the around mean-field formula \cite{Czyzyk_LDA+U_AMF_suppl} for the double-counting correlation, and we crosschecked results with the self-interaction correction formula \cite{Anisimov_LDA+U_SIC_suppl, Liechtenstein_LDA+U_suppl}.

\begin{figure}[ht]
\begin{center}
\includegraphics[width=1.0\columnwidth]{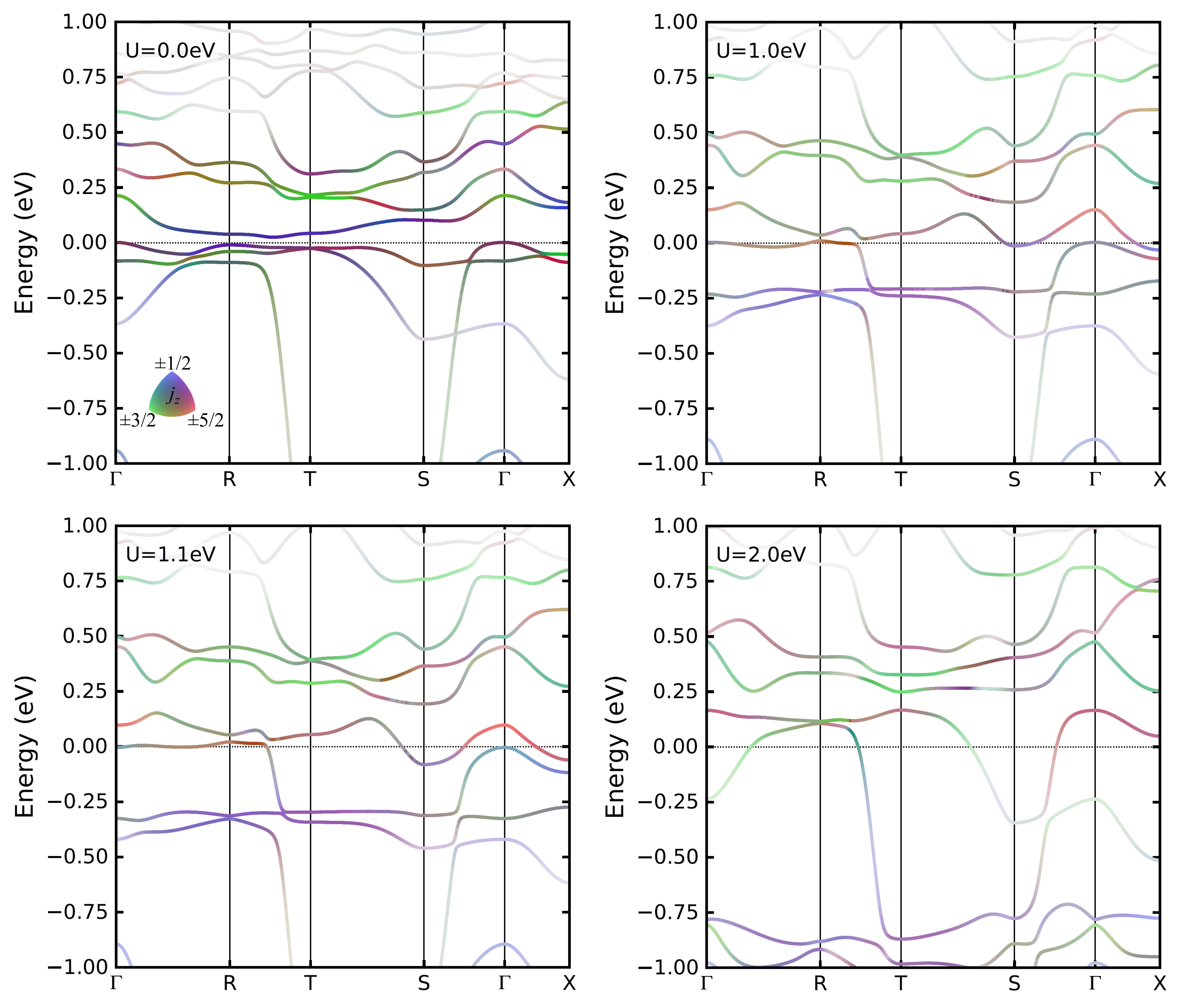}	
\caption{
The calculated band structure by the DFT and GGA$+U$ along the high symmetry line. Compared to the DFT results, two $j=5/2$ bands are pushed downward in energy about $U/2$, leaving the band structure near the Fermi level dominated by the $5f$ $j_z=\pm5/2$ and $\pm3/2$ states. The blue, green, red, and grey colors represent the weight of orbitals for $j_z=\pm1/2$, $\pm3/2$, $\pm5/2$, and other components, respectively.
\label{fig:band}}
\end{center}
\end{figure}

\begin{figure}[ht]
\begin{center}
\includegraphics[width=1.0\columnwidth]{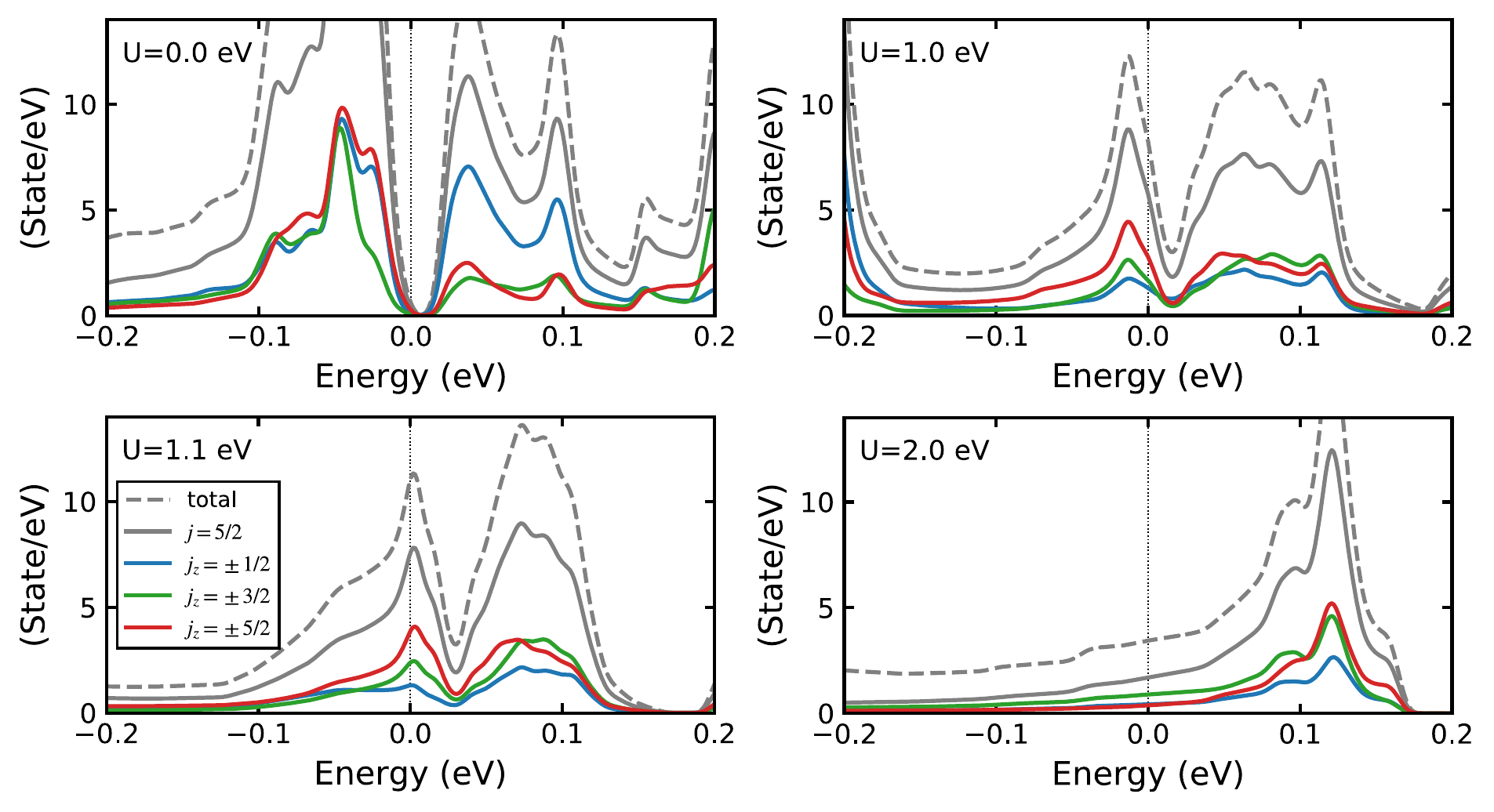}	
\caption{
Partial density of states for U-5$f$ orbitals $j=5/2$, $j_z=\pm1/2$, $\pm3/2$, and $\pm5/2$ near the Fermi level. Dashed line represents the total density of states. 
\label{fig:dos}}
\end{center}
\end{figure}

Figure \ref{fig:band} shows the band structure by DFT and GGA$+U$ along high symmetry lines.
Uranium $5f$ orbitals split into the $j=5/2$ and $j=7/2$ manifolds due to the spin-orbit coupling.
We observe near the Fermi level doubly degenerated bands of the $j=5/2$ multiplet, two of which are valence bands, and other four bands are conduction bands.
The $j=7/2$ states are located at around $1.0$ eV.
The electron occupancy of $5f$ orbitals is around $n_f\sim2.5$, which is larger than the nominal occupancy of U$^{2+}$ ion $n^0_f=2$.
Note that the dispersive bands at around $-0.5$ eV are composed of $5f$ orbitals mixed with U-$d_{3z^2-r^2}$ orbital and Te(2)-$p_y$ orbital.
The DFT band structure is consistent with the previous result \cite{Aoki_UTe2_suppl}.
Moreover, in Fig.~\ref{fig:band}, $j_z$ orbital characters are highlighted.
The $j_z=\pm1/2$, $\pm3/2$, and $\pm5/2$ states are quite entangled.
The flat bands near the band gap are dominated by $j_z=\pm1/2$ and $\pm5/2$ states in the DFT calculation.
The band structures by GGA$+U$ are shown for $U=1.0$, $1.1$, and $2.0$ eV.
Compared to the DFT results, two $j=5/2$ bands are shifted downward in energy about $U/2$, leaving the band structure near the Fermi level dominated by the $5f$ $j_z=\pm5/2$ and $\pm3/2$ states.
We can see an electron pocket at the $S$ and $X$ points and a hole pocket at the $R$ point.
Figure \ref{fig:dos} shows partial density of states for U-5$f$ orbitals $j=5/2$, $j_z=\pm1/2$, $\pm3/2$, and $\pm5/2$ near the Fermi level.
We can confirm the insulator-metal transition.
The $j_z=\pm5/2$ states are dominated at the Fermi level for $U=1.0$ and $1.1$ eV.

\begin{table}[h]
\caption{Atomic coordinates of UTe$_2$. Lattice constants are $a=4.1617$, $b=6.1276$, and $c=13.9650$ \AA. }
\label{tb:struct}
\begin{tabular*}{0.5\columnwidth}{@{\extracolsep{\fill}}llll}
\hline\hline
Atom & $x$ & $y$ & $z$ \\ \hline
U ($4i$) & $0$ & $0$ & $0.13480$ \\
Te1 ($4j$) & $0.5$ & $0$ & $0.29770$ \\
Te2 ($4h$) & $0$ & $0.2510$ & $0.5$ \\
\hline\hline 
\end{tabular*}
\end{table}

\clearpage
\section{Topological classification of superconducting gap structures}
Table~IV in the main text gives a classification of superconducting \textit{order parameter}, which enables us to speculate gap structures of UTe$_2$.
However, it is desirable to use the classification of \textit{gap structures} in terms of symmetry and topology~\cite{Yarzhemsky1992_suppl, Norman1995_suppl, Micklitz2009_suppl, Nomoto2017_suppl, Micklitz2017_PRL_suppl, Sumita2017_suppl, Sumita2018_suppl, Kobayashi2014_suppl, Kobayashi2016_suppl, Kobayashi2018_suppl, Sumita2019_suppl} because symmetry-protected gap nodes are precisely obtained.
This scheme focuses on high-symmetry points in the Brillouin zone in Fig.~2(a) of the main text: mirror planes ($k_z = 0, \, 2\pi / c$, $k_y = 0$, and $k_x = 0$) and rotational axes labeled by $\Lambda$, $\Delta$, $U$, $\Sigma$, and $F$.
Crystal symmetry allowed on each high-symmetry $\bm{k}$ point is represented by a unitary little cogroup $\bar{\cal G}^{\bm{k}}$, whose irreducible representation is denoted by $\alpha$.

At zero magnetic field, we define time-reversal symmetry (TRS), particle-hole symmetry (PHS), and chiral symmetry (CS) operators preserving any $\bm{k}$ points by $\mathfrak{T} \equiv {\cal T I}$, $\mathfrak{C} \equiv {\cal C I}$, and $\Gamma \equiv {\cal T C}$, respectively.
Thus the intrinsic symmetry is represented by the following group:
\begin{equation}
 \bar{\mathfrak{G}}^{\bm{k}} = \bar{\cal G}^{\bm{k}} + \mathfrak{T} \bar{\cal G}^{\bm{k}} + \mathfrak{C} \bar{\cal G}^{\bm{k}} + \Gamma \bar{\cal G}^{\bm{k}}.
\end{equation}
Then, using the factor system $\{z_{g, h}^{\bm{k}}\} \in Z^2(\bar{\mathfrak{G}}^{\bm{k}}, \mathrm{U}(1)_\phi)$, we execute the \textit{Wigner criteria}~\cite{Wigner_suppl, Herring1937_suppl, Inui-Tanabe-Onodera_suppl, Bradley_suppl, Shiozaki2018_arXiv_suppl} for $\mathfrak{T}$ and $\mathfrak{C}$, and the \textit{orthogonality test}~\cite{Inui-Tanabe-Onodera_suppl, Shiozaki2018_arXiv_suppl} for $\Gamma$:
\begin{align}
 W_\alpha^{\mathfrak{T}} &\equiv \frac{1}{|\bar{\cal G}^{\bm{k}}|} \sum_{g \in \bar{\cal G}^{\bm{k}}} z_{\mathfrak{T} g, \mathfrak{T} g}^{\bm{k}} \chi[\bar{\gamma}^{\bm{k}}_\alpha((\mathfrak{T} g)^2)] =
 \begin{cases}
  1, \\
  -1, \\
  0,
 \end{cases} \label{eq:Wigner_criterion_T} \\
 W_\alpha^{\mathfrak{C}} &\equiv \frac{1}{|\bar{\cal G}^{\bm{k}}|} \sum_{g \in \bar{\cal G}^{\bm{k}}} z_{\mathfrak{C} g, \mathfrak{C} g}^{\bm{k}} \chi[\bar{\gamma}^{\bm{k}}_\alpha((\mathfrak{C} g)^2)] =
 \begin{cases}
  1, \\
  -1, \\
  0,
 \end{cases} \label{eq:Wigner_criterion_C} \\
 W_\alpha^\Gamma &\equiv \frac{1}{|\bar{\cal G}^{\bm{k}}|} \sum_{g \in \bar{\cal G}^{\bm{k}}} \frac{z_{g, \Gamma}^{\bm{k} *}}{z_{\Gamma, \Gamma^{-1} g \Gamma}^{\bm{k} *}} \chi[\bar{\gamma}^{\bm{k}}_\alpha(\Gamma^{-1} g \Gamma)^*] \chi[\bar{\gamma}^{\bm{k}}_\alpha(g)] =
 \begin{cases}
  1, \\
  0.
 \end{cases} \label{eq:orthogonality_test_G}
\end{align}
From Eqs.~\eqref{eq:Wigner_criterion_T}-\eqref{eq:orthogonality_test_G}, we obtain the set of $(W_\alpha^{\mathfrak{T}}, W_\alpha^{\mathfrak{C}}, W_\alpha^\Gamma)$, which indicates the \textit{effective} Altland-Zirnbauer (EAZ) symmetry class of the Bogoliubov-de Gennes Hamiltonian on the high-symmetry point~\cite{Altland1997_suppl, Shiozaki2018_arXiv_suppl}.
The EAZ class gives us classification of gap structures, assuming that the high-symmetry planes or lines intersect normal-state Fermi surfaces (for details, see Ref.~\cite{Sumita2019_suppl}).
After some algebra, we obtain the gap classification shown in Table~\ref{tab:classification_gap}(a).
%Note that the magnetic space group is $Immm1'$ in zero magnetic field, while it is reduced to $Im'm'm$ under $b$ magnetic field.

\begin{table}[tbp]
 \centering
 \caption{Classification of superconducting gap structures on mirror planes and rotational axes under (a) zero magnetic field and (b) magnetic fields along the $b$ axis. In the tables, we show the EAZ class for each pairing symmetry. Letters in parenthesis represent the absence ($0$) or presence ($\mathbb{Z}$ or $\mathbb{Z}_2$) of a topological invariant, and gap structures: (G) full gap, (P) point nodes, or (L) line nodes.}
 \label{tab:classification_gap}
  \begin{tabular}[t]{lcccccc}
  \multicolumn{7}{c}{(a) Zero magnetic field} \\ \hline\hline
  $D_{2h}$ & $k_z = 0, \, 2\pi / c$ & $k_y = 0$ & $k_x = 0$ & $k_z$ axis ($\Lambda$) & $k_y$ axes ($\Delta, U$) & $k_x$ axes ($\Sigma, F$) \\ \hline
  $A_u$    & C    ($0$, G) & C    ($0$, G) & C    ($0$, G) & CI  ($0$, G)            & CI  ($0$, G)            & CI  ($0$, G)            \\
  $B_{1u}$ & C    ($0$, G) & AIII ($0$, G) & AIII ($0$, G) & BDI ($\mathbb{Z}_2$, P) & CI  ($0$, G)            & CI  ($0$, G)            \\
  $B_{2u}$ & AIII ($0$, G) & C    ($0$, G) & AIII ($0$, G) & CI  ($0$, G)            & BDI ($\mathbb{Z}_2$, P) & CI  ($0$, G)            \\
  $B_{3u}$ & AIII ($0$, G) & AIII ($0$, G) & C    ($0$, G) & CI  ($0$, G)            & CI  ($0$, G)            & BDI ($\mathbb{Z}_2$, P) \\ \hline\hline
 \end{tabular}
 \hspace{10mm}
 \begin{tabular}[t]{lcc}
  \multicolumn{3}{c}{(b) Magnetic field $\parallel b$} \\ \hline\hline
  $C_{2h}^y$ & $k_y = 0$ & $k_y$ axes ($\Delta, U$) \\ \hline
  $A_u$ & C (0, G)            & AI ($\mathbb{Z}$, P) \\
  $B_u$ & A ($\mathbb{Z}$, L) & CI (0, G)            \\ \hline\hline
 \end{tabular}
\end{table}

Under magnetic field along the $b$ axis, on the other hand, the intrinsic symmetry is changed.
On the mirror plane ($k_y = 0$), TRS $\mathfrak{T}$ and CS $\Gamma$ are ill-defined, while PHS $\mathfrak{C}$ is introduced by ${\cal C I}$.
On the rotational axes ($\Delta, U$), we define TRS, PHS, and CS by $\mathfrak{T} \equiv {\cal T} C_{2z}$, $\mathfrak{C} \equiv {\cal C I}$, and $\Gamma \equiv {\cal T} C_{2z} {\cal C I}$, respectively.
Eqs.~\eqref{eq:Wigner_criterion_T}-\eqref{eq:orthogonality_test_G} are also applied into the cases, which result in Table~\ref{tab:classification_gap}(b).
Note that $W_\alpha^{\mathfrak{T}} = W_\alpha^{\Gamma} = 0$ on the mirror plane since the little cogroup $\bar{\mathfrak{G}}^{\bm{k}}$ has \textit{no} TRS and CS.

\end{document}